\documentclass[sigconf]{acmart}
\AtBeginDocument{%
  \providecommand\BibTeX{{%
    \normalfont B\kern-0.5em{\scshape i\kern-0.25em b}\kern-0.8em\TeX}}}

\usepackage{multirow}
\usepackage{float} 
\usepackage{amsmath}
\usepackage[ruled,linesnumbered]{algorithm2e}
\usepackage{graphicx}
\usepackage{subfigure}
\usepackage{epstopdf}
\usepackage{tablefootnote}
\usepackage[flushleft]{threeparttable}
\usepackage{diagbox}
\usepackage{booktabs}
\usepackage{siunitx}
\usepackage{footmisc}
\usepackage{footnote}
\usepackage{enumitem}
\usepackage[normalem]{ulem}
\useunder{\uline}{\ul}{}

\begin{document}

\title{DELTA: Pre-train a Discriminative Encoder for Legal Case Retrieval via Structural Word Alignment}

\author{Haitao Li}
\affiliation{DCST, Tsinghua University}
\affiliation{Quan Cheng Laboratory}
\email{liht22@mails.tsinghua.edu.cn}

\author{Qingyao Ai}
\affiliation{DCST, Tsinghua University}
\authornote{Corresponding author}
\affiliation{Quan Cheng Laboratory}
\email{aiqy@tsinghua.edu.cn}

\author{Xinyan Han}
\affiliation{DCST, Tsinghua University}
\affiliation{Quan Cheng Laboratory}
\email{hanxinyan20@gmail.com}

\author{Jia Chen}
\affiliation{DCST, Tsinghua University}
\affiliation{Quan Cheng Laboratory}
\email{chenjia0831@gmail.com}

\author{Qian Dong}
\affiliation{DCST, Tsinghua University}
\affiliation{Quan Cheng Laboratory}
\email{dq22@mails.tsinghua.edu.cn}


\author{Yiqun Liu}

\affiliation{DCST, Tsinghua University}
\affiliation{Zhongguancun Laboratory}
\email{yiqunliu@tsinghua.edu.cn}

\author{Chong Chen}
\affiliation{Huawei Cloud BU}
\email{chenchong55@huawei.com}

\author{Qi Tian}
\affiliation{Huawei Cloud BU}
\email{tian.qi1@huawei.com}

\begin{abstract}
Recent research demonstrates the effectiveness of using pre-trained language models for legal case retrieval. 
Most of the existing works focus on improving the representation ability for the contextualized embedding of the $[CLS]$ token and calculate relevance using textual semantic similarity.
However, in the legal domain, textual semantic similarity does not always imply that the cases are relevant enough. Instead, relevance in legal cases primarily depends on the similarity of key facts that impact the final judgment. Without proper treatments, the discriminative ability of learned representations could be limited since legal cases are lengthy and contain numerous non-key facts.

To this end, we introduce DELTA, a discriminative model designed for legal case retrieval.
The basic idea involves pinpointing key facts in legal cases and pulling the contextualized embedding of the $[CLS]$ token closer to the key facts while pushing away from the non-key facts, which can warm up the case embedding space in an unsupervised manner. To be specific, this study brings the word alignment mechanism to the contextual masked auto-encoder. First, we leverage shallow decoders to create information bottlenecks, aiming to enhance the representation ability. Second, we employ the deep decoder to enable ``translation'' between different structures, with the goal of pinpointing key facts to enhance discriminative ability. Comprehensive experiments conducted on publicly available legal benchmarks show that our approach can outperform existing state-of-the-art methods in legal case retrieval.
It provides a new perspective on the in-depth understanding and processing of legal case documents.
\end{abstract}

\maketitle

\section{Introduction}
Legal case retrieval, which focuses on retrieving relevant cases for a query case, is essential for supporting legal reasoning and decision-making. In judicial practice, legal practitioners need to consult and review applicable precedents before making a judgment. Recently, with the increasing number of digitized legal cases, legal professionals face the challenge of effectively and efficiently finding relevant information from the vast amount of legal literature. The development and application of legal case retrieval technology have become the key to solving this problem~\cite{shao2020bert,SAILER,li2023thuircoliee,xie2023t2ranking}.

Apart from using classic keyword matching models to retrieve relevant legal case documents, neural retrieval models constructed with pre-trained language models, which are capable of capturing the latent semantics of text documents, have attracted significant attention in the field of legal case retrieval~\cite{li2023thuircoliee,bhattacharya2022legal,su2024caseformer}. 
Several studies have tried to apply the out-of-the-box pre-trained language models such as BERT and RoBERTa to build semantic representations for legal case documents and retrieve relevant cases based on their representation similarities~\cite{bhattacharya2022legal,li2023thuircoliee}. More studies have focused on how to tailor the pre-trained representation models for legal case retrieval by designing new model structures and formulating specific training objectives~\cite{xiao2021lawformer,SAILER,su2024caseformer}. For instance, Xiao et al.~\cite{xiao2021lawformer} introduced Lawformer, designed to process thousands of tokens, thereby tackling the challenge posed by the lengthy nature of legal documents. Similarly, Li et al.~\cite{SAILER} developed a structure-aware framework named SAILER, which fully utilizes structural information in legal cases to better pretrain the text encoder. These models, crafted for legal contexts, demonstrate superior performance on multiple public legal case retrieval benchmarks.

\begin{figure}[t]
\centering
\includegraphics[width=\columnwidth]{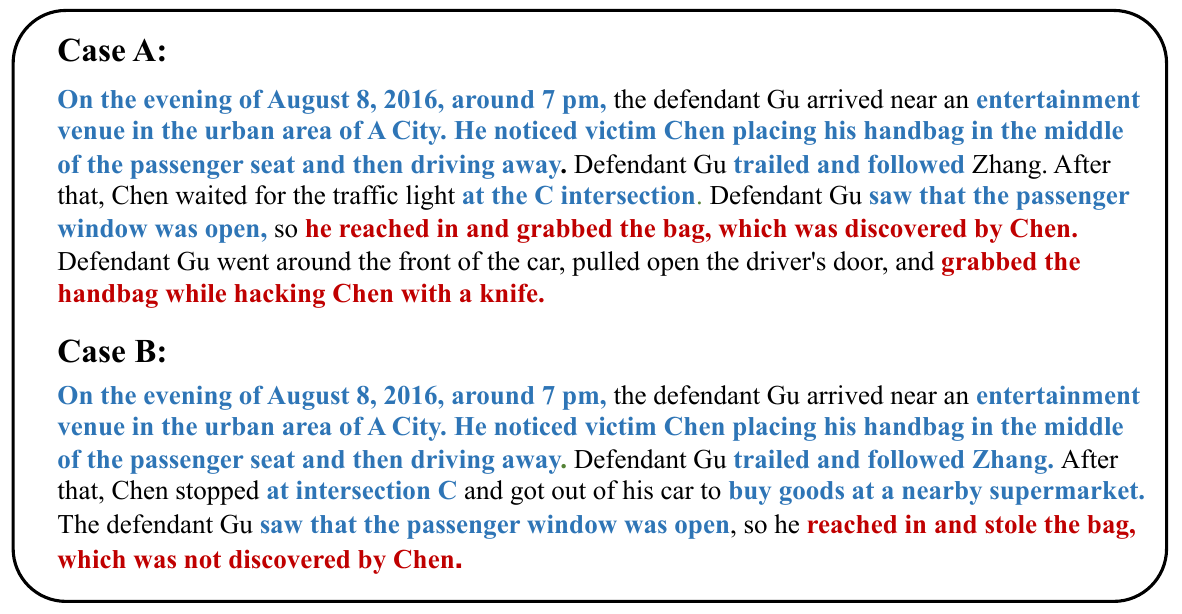}
\vspace{-4mm}
\caption{An illustrative example of relevance in legal case retrieval. The key facts in these cases are highlighted in red and the non-key facts in blue. Despite sharing a substantial number of words, these two cases are legally irrelevant. The divergence in key facts results in distinctly different judgments.}
\label{case}
\vspace{-6mm}
\end{figure}

Despite their success, existing pre-trained retrieval models for legal case retrieval are far from being perfect due to several problems. One particularly significant issue is their incapability of discriminating key legal facts,  which are crucial for determining case relevance, from other texts that simply describe case background and unimportant facts.  
As highlighted in prior research~\cite{ma2021lecard,shaotois,li2023lecardv2}, the relevance determination in legal case retrieval is complex and fundamentally different from general search relevance. This is because, unlike open-domain retrieval which mostly relies on keyword matching or text semantic similarity, legal case relevance focuses more on the identification of key facts that are crucial for case decisions. 
For instance, Figure ~\ref{case} shows that two cases, even with high textual similarity, can be legally irrelevant due to their differences in key facts. To illustrate, Case A concerns a robbery, whereas Case B pertains to theft, each leading to a completely different judgment.
In the pursuit of relevant cases, legal practitioners give precedence to key facts rather than non-key facts such as background information.
Existing pre-trained models focus on constructing better representations that capture the text semantics of individual case documents.
However, in the legal domain, better semantic representation vectors do not always lead to better discrimination of legal relevance if the representations focus on capturing facts that are unimportant from legal perspectives.

To this end, in this paper, we propose a novel framework called DELTA, which stands for Pre-training a \textbf{D}iscriminative \textbf{E}ncoder for \textbf{L}egal Case Re\textbf{T}rieval via Structural Word \textbf{A}lignment. Specifically, DELTA employs an encoder-decoder architecture to achieve an in-depth understanding of legal cases and effective extraction of key information.
Besides employing shallow decoders to create a bottleneck for the $[CLS]$ token and generate high-quality textual representations, DELTA also incorporates the Structural Word Alignment (SWA) task to identify key facts by ``translating" the Fact section of a legal case document to the legal analysis and decisions in the Reasoning section. 
Furthermore, DELTA enhances the alignment between the case representation and its key factual information in the semantic space.
This alignment is achieved by pulling the case representation closer to the key facts while simultaneously pushing it away from background information within the case document.
The whole algorithm not only enhances the discriminative ability of the representation models from the legal perspective but also helps legal users better track the key facts in retrieved cases, thus making the results more interpretable and trustworthy.

To validate the effectiveness of our model, we carried out comprehensive experiments on Chinese and English legal benchmarks in both zero-shot and fine-tuning settings. Empirical results indicate that DELTA significantly outperforms current state-of-the-art baselines. The main contributions of this paper are summarised as follows: 

\begin{enumerate} 
\item We propose a novel pre-training framework DELTA tailored for legal case retrieval. By combining the deep decoder and shallow decoders, DELTA can enhance the discriminative ability of the case encoder from a legal perspective. 
\item We introduce the Structural Word Alignment (SWA) task to identify key facts in legal cases. To the best of our knowledge, DELTA is the first work to apply word alignment in legal case retrieval, which provides new perspectives for in-depth understanding and processing of legal cases.	 
\item Extensive experiments are conducted on publicly available Chinese and English legal benchmarks. These experiments illustrate that DELTA achieves significant improvements over other competing retrievers in legal case retrieval.
\end{enumerate}

\section{Related Work}


\subsection{Legal Case Retrieval}
Legal case retrieval, a specialized task in information retrieval, has attracted considerable attention from both academia and industry.
Researchers in legal case retrieval primarily focus on two categories of models: Expert Knowledge-based models~\cite{zeng2005knowledge,saravanan2009improving} and Natural Language Processing (NLP) models~\cite{shao2020bert,xiao2021lawformer,chalkidis2020legal,SAILER,li2023thuir2}. Expert knowledge-based modeling enhances case representation by extending legal elements~\cite{zeng2005knowledge} or developing ontological frameworks~\cite{saravanan2009improving}, including the introduction of new sub-elements to the legal problem element for a more comprehensive representation of legal cases. On the other hand, Natural Language Processing models employ deep learning techniques such as BERT and its variants to capture semantic similarities between cases at the text level~\cite{dong2023i3}. These models show great potential for legal case retrieval.

Recently, numerous studies have addressed the challenge of long texts in the legal field by segmenting the text into paragraphs or enlarging the inputs for language models. 
For instance, Shao et al.~\cite{shao2020bert} segment legal documents into multiple paragraphs and subsequently aggregate the scores. Xiao et al.~\cite{xiao2021lawformer} introduced Lawformer, which increases the model's text input length by enhancing the attention mechanism.
Furthermore, many researchers have attempted to enhance performance by pre-training models on extensive legal corpora. For instance, the development of LEGAL-BERT~\cite{chalkidis2020legal} involved collecting a significant array of English legal texts from various domains for pre-training. Li et al.~\cite{SAILER} employed the specific structure of legal documents to pre-train SAILER, achieving state-of-the-art results on the legal dataset. Despite great success, the primary objective of these models is to enhance representing capabilities. However, in the legal domain, semantic similarity does not always correspond to case relevance. In this paper, our objective is to identify key facts within legal documents, thereby enhancing the discriminative ability of representation vectors.

\subsection{Word Alignment in Language Translation}

Word Alignment (WA) is a crucial component of statistical machine translation, primarily focused on establishing correspondences between words in a sentence pair~\cite{li2019word}. This task plays a critical role in comprehending the relationships between languages and facilitating cross-language translation. In traditional statistical machine translation~\cite{dyer2013simple,och2003systematic}, word alignment usually requires an annotated parallel corpus, i.e., correspondences need to be manually created for each word in a sentence. This procedure is both labor-intensive and costly.

With the rapid development of Neural Machine Translation, alignment by attention brings a more flexible and efficient solution~\cite{bahdanau2014neural,chatterjee2017guiding,li2019word,garg2019jointly,zenkel2019adding,liu2016neural}. Neural machine translation can automatically learn word alignment from a corpus without manual annotation. Bahdanau et al.~\cite{bahdanau2014neural} were the first to demonstrate an example of word alignment using attention mechanisms in the RNNSearch model. Subsequently, Liu et al. ~\cite{liu2016neural} improved attention with the annotation results obtained by the statistical alignment tool, expecting better alignment results.
Grag et al.~\cite{garg2019jointly} showed that the penultimate layer of the transformer tends to learn better alignment and designed multitask learning to jointly learn translation and word alignment.
Inspired by this work, we attempt to utilize word alignment between different structures to identify key facts in legal cases.

\section{Preliminary}

In this section, we introduce the task description, dual encoder architecture and structure in legal case documents.

\begin{figure*}[t]
\centering
\includegraphics[width=0.8\textwidth]{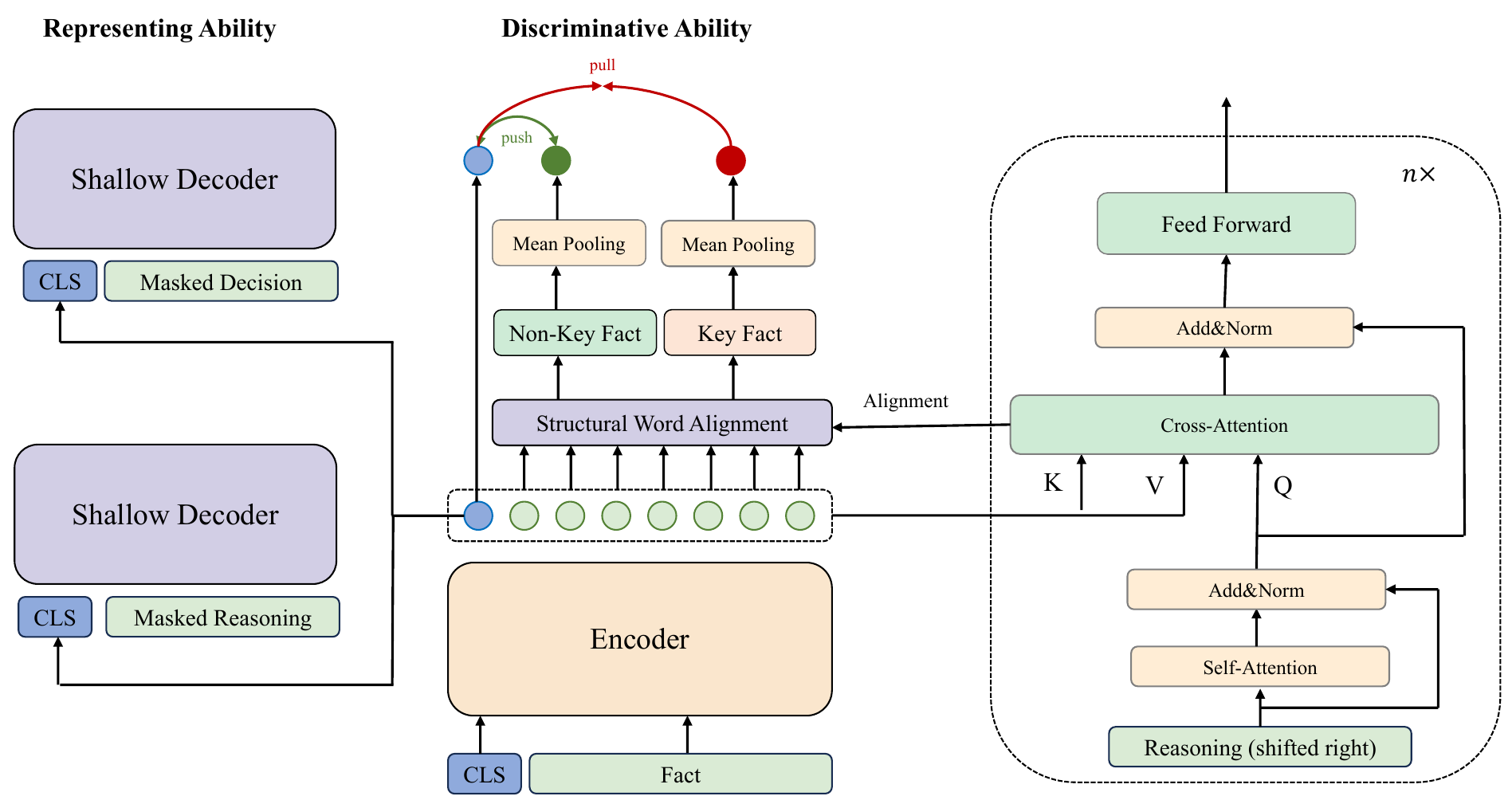}
\caption{Pre-training designs of DELTA. DELTA creates information bottlenecks with two shallow decoders to improve representing ability of $[CLS]$ vector. Furthermore, Structural Word Alignment task is employed to identify key facts. DELTA pulls $[CLS]$ vectors closer to key facts and pushes them away from the non-key facts to enhance discriminative ability.}
\label{model}
\end{figure*}

\subsection{Task Description}
The main purpose of legal case retrieval is to identify the relevant cases from the candidate cases for each query case. Formally, for a query $q$, the legal practitioner needs to find the top-$k$ relevant cases from the candidate set $\mathbf{C}$.
In the legal domain, these relevant cases $\mathbf{C}_q = c_{1}^*, c_{2}^*,......, c_{k}^*$ are known as precedents, referring to historical cases that provide support for the judgment of the query case.

In most judicial practices, query cases usually require further discussion and analysis to make a final judgment, thus they only include the basic facts of the case. Candidate cases, on the other hand, are legal case documents that have already been definitively adjudicated historically, containing the complete structure.
Based on the above setting, this paper assumes that the query cases are primarily concerned with the basic facts in the legal case documents. Such assumptions can help us to more accurately model and analyze the actual situations that legal professionals face when conducting legal case retrieval in reality.

\subsection{Dual Encoder}

In this paper, we focus on the dual-encoder architecture to achieve efficient retrieval, which has been widely used in several retrieval tasks~\cite{dongi3,li2023constructing,lu2021less,wang2022simlm}. The dual encoder separately maps the query and the document to low-dimensional representations. This low-dimensional representation is generally called the dense vector and is trained to capture the semantics of the input text. Then, the relevance score is calculated using simple functions such as cosine and dot products.
Following common practice, we typically utilize the hidden vector of $[CLS]$ token from the last hidden layer as the dense representation. 
Specifically, given a case $c_i=([CLS],w_1,...,w_n)$, where $w_k$ represents ordinary token and $n$ is the length of this case. Then, the case is represented as follows:
$$
h_{c_i} = h_{[CLS]} = Encoder(c_i)
$$
In general, high-quality representation vectors possess the ability not only to accurately capture textual semantics but also to discriminate relevant documents from irrelevant ones.
Previous work~\cite{lu2021less,wang2022simlm,wu2022contextual,liu2022retromae} focused on creating information bottlenecks to force $[CLS]$ token into better representing text, while ignoring discriminative ability. 


\subsection{Structure in Legal Cases}

Whether in countries with the Case Law system or the Statute Law system, legal case documents typically have clear structures and intrinsic writing logics~\cite{SAILER}. Specifically, it is common for a complete legal case to contain three parts: Fact, Reasoning, and Decision. The Fact section focuses on the argument, evidence, and basic facts. Since arguments and evidence are not all useful, Fact section usually contains a great deal of non-key facts.
The Reasoning section reveals how the court selects and applies the legal rules. The Decision section is the definitive response of the court to the legal dispute.

It is worth noting that these three sections do not exist independently. The key facts in the Facts section will be carefully analyzed in the Reasoning section and further influence the final decision. In other words, the Reasoning section contains all the key facts of the case, which is essential to determining the relevance.
In practice, when legal practitioners draft a legal case document, they first analyze the key facts and subsequently construct the Reasoning section based on their experience.
This process can be regarded as ``translating'' the Fact section into the Reasoning section. Numerous words or phrases within these two sections exhibit correspondences. For example, when the Fact section describes an act as ``stealing a cell phone while someone is unaware of it'', the legal practitioner may categorize it as a ``secret theft'' in the Reasoning section. 
Similarly, the term ``particularly large amounts'' in the Reasoning section typically indicates that the amount in fact exceeds three million RMB. Understanding the structures in legal cases and learning the correspondence between structures is essential to improve the performance of legal case retrieval.
 
\section{Method}
In this section, we present the framework and training methods of DELTA in detail.

\subsection{Overview}

The framework of DELTA is shown as Figure ~\ref{model}. DELTA consists of three components, i.e., fact encoder, shallow decoders for different case sections, and a deep decoder for word alignment. The unified encoder maps the Fact section into a dense embedding. The optimization goal is to make this dense embedding more expressive and distinguishable. To achieve this objective, we develop the following strategies.

\begin{enumerate} 
\item \textbf{Represention Ability:} 
Shallow decoders are employed to create information bottlenecks, compelling the encoder to provide better text representations. Specifically, the Reasoning and Decision sections are aggressively masked and combined with the representation vectors of the Fact section to reconstruct the text. Due to the restricted capacity of the shallow decoders, the encoder is forced to better represent the input text.
\item \textbf{Discriminative Ability:} DELTA improves the discriminative ability of the representation embeddings through the Structural Word Alignment (SWA) task. The intuitive idea is that $[CLS]$ embedding should be close to the representation of key facts and pull away from non-key facts. Hence, we employ a deep decoder to learn the ``translation'' from the Fact to the Reasoning section. After that, the attention matrix is utilized to identify the key facts within the Fact section. 
\end{enumerate}


\subsection{Fact Encoder}
The Fact Encoder encodes the Fact section into a high-quality representation vector to perform effective retrieval.
In particular, given a Fact section $F = [f_1,...,f_n]$, we randomly sample and replace part of its tokens with $[MASK]$ token. To preserve enough information, only a small portion (<=15\%) of the token is replaced. We define the masked tokens as $m(F)$ and input the masked fact $F_{mask}$ to the encoder $\Psi_F$.
After that, we can get the contextualized embeddings of $[CLS]$ token $h_F$ and other ordinary tokens $H_F$:
\begin{equation}\label{eqn-19} 
  h_F,H_F = \Psi_F(F_{mask}) 
\end{equation}

As discussed in previous work~\cite{devlin2018bert,liu2019roberta}, the Masked Language Modeling (MLM) task can contribute to better contextual representation. Therefore, we include MLM as one of the pre-training objectives. The loss function of MLM is shown as follows:

\begin{equation}\label{eqn-3} 
  L_{MLM} = -\sum_{x^{'}\in m(F)}\log p(x^{'} | F \backslash m(F))
\end{equation}

\subsection{Shallow Decoder}
Following the setting of SAILER~\cite{SAILER}, we introduce two shallow decoders i.e., Reasoning Decoder and Decision Decoder, to model the structure in legal cases. In this process, we concatenate the dense vector $h_F$ with contextual sentence embedding and feed it into shallow decoders for text reconstruction.

Specifically, we select some original tokens from the Reasoning section $R = [r_1,r_2,...,r_n]$ and the Decision section $D = [d_1,d_2,...,d_n]$ and replace them with $[MASK]$ token. Aggressive masking rates (>=30\%) are employed to provide adequate difficulty in reconstruction. 
Then, the original $[CLS]$ vectors in the Reasoning and Decision section are replaced with the dense representation $h_F$ from Fact Encoder. Formally, the input sequence to decoders are as follows:
\begin{equation}\label{eqn-5} 
  INPUT_{\rm reasoning} = [h_F, \boldsymbol{r_1},...,\boldsymbol{r_n}];
\end{equation}
\begin{equation}\label{eqn-6} 
  INPUT_{\rm decision} = [h_F, \boldsymbol{d_1},...,\boldsymbol{d_n}];
\end{equation}
where $\boldsymbol{r_i}$ and $\boldsymbol{d_i}$ denote the embeddings of token $r_i$ and $d_i$. We define the masked tokens as $m(R)$ and $m(D)$. The training objective of shallow decoders can be defined as:

\begin{equation}\label{eqn-7} 
  L_{REA} = -\sum_{x^{'}\in m(R)}\log p(x^{'} | [h_{F},R \backslash m(R)])
\end{equation}

\begin{equation}\label{eqn-8} 
  L_{DEC} = -\sum_{x^{'}\in m(D) }\log p(x^{'} | [h_{F},D \backslash m(D))])
\end{equation}

As discussed in previous work~\cite{lu2021less}, the shallow decoder has limited capabilities, so $h_f$ is forced to represent more information for text reconstruction. For a detailed masked method, please refer to SAILER~\cite{SAILER}.

\subsection{Structural Word Alignment}
As mentioned before, there are numerous non-key facts in legal case documents that do not significantly contribute to determining case relevance.
Since the Reasoning section discusses and analyzes all the key facts, each of which has a corresponding description, we attempt to identify the key fact tokens in the Fact section through the word alignment mechanism.
However, annotated word alignment is challenging and labor-intensive. Instead, there has been some work exploring unsupervised word alignment with neural machine translation, which has proven to be effective~\cite{garg2019jointly,zenkel2019adding}. Inspired by these studies, we propose the Structural Word Alignment (SWA) task in this section.

Specifically, given a Fact section $F = [f_1,...,f_I]$ that serves as the source sentence and a Reasoning section $R = [r_1,r_2,...,r_J]$ as the target sentence, where $I$ and $J$ represent the lengths of Fact and Reasoning section respectively, an alignment $\Lambda$ is defined as a subset of the Cartesian product of the word positions~\cite{och2003systematic}:
$$
\Lambda \subseteq \{(i,j):i=1,...,I;j=1,...,J \} 
$$

The goal of the word alignment task is to establish a discrete alignment, representing a many-to-many mapping of source words to their corresponding translations in the target sentence. To achieve unsupervised word alignment, DELTA introduces another decoder, with sufficiently deep layers to effectively perform the translation process. 
This decoder is autoregressive and its training objective can be formulated as:
\begin{equation}\label{eqn-1} 
  L_{TLM} = -\sum_{j=1}^{J}\log p(r_j | F, r_{<j},\theta)
\end{equation}
where $\theta$ denotes the parameters of the deep decoder. Different from the shallow decoder, all ordinary tokens in the Fact section contribute to decoding the next token.
It is also worth noting that this loss function does not optimize the parameters of the encoder to preserve its original encoding capability.

In this paper, we focus on guiding the cross-attention sub-layer in the decoder to obtain the alignment relation.
Formally, the representation of the $i^{th}$ target token in the decoder as $q^i \in \mathbb{R}^{1*d_{emb}}$, where $d_{emb}$ denotes the dimension of vectors. 
The output vectors of all source tokens from the encoder serve as the key matrix $K \in \mathbb{R}^{I*d_{emb}}$ and the value matrix $V \in \mathbb{R}^{I*d_{emb}}$.  It is worth noting that the $K$ and $V$ matrices are identical, which corresponds to the outputs of the encoder. Then, the $N$ heads project the query vector and the key and value matrices into distinct subspaces:
\begin{equation}\label{eqn-9} 
\tilde{q}_n^i = q^iW_n^Q, \tilde{K}_n=KW_n^K, \tilde{V}_n=VW_n^V  
\end{equation}
\begin{equation}\label{eqn-10} 
H_n^i = Attention(\tilde{q}_n^i,\tilde {K}_n,\tilde {V}_n)
\end{equation}
\begin{equation}\label{eqn-11} 
M(q^i,K,V) = Concat(H_1^i,...,H_N^i)W^O
\end{equation}
where $W_n^Q$, $W_n^K$, $W_n^V$ and $W^O$ are all trainable parameters of the $n^{th}$ head. $Concat(\cdot)$ denotes the concatenate of multi-head attention. The scaled dot-product attention is employed by each head:
\begin{equation}\label{eqn-12} 
Attention(\tilde {q}_n^i,\tilde{K}_n^i,\tilde {V}_n^i) = a_n^i\tilde {V}_n
\end{equation}

\begin{equation}\label{eqn-13} 
a_n^i = softmax(\frac{\tilde{q}_n^i\tilde{K}_n^T}{\sqrt{d_{emb}} } )
\end{equation}
where $a_n^i$ represents the attention probabilities for the $i^{th}$ target token across all source tokens in the $n^{th}$ attention head. The multi-head attention mechanism of the transformer generates multiple attention matrices. To gain a deeper insight into the behavior of the encoder-decoder attention, we average the attention matrices across all heads within each layer to get $a^i$.
The word alignment $\Lambda$ can be readily extracted from the attention weight $a^i$ according to the style of maximum a posterior strategy (MAP):
\begin{equation}\label{eqn-4} 
  	\Lambda_{i,j}=\left\{
		\begin{aligned}
			1 & & j = \text{arg} \max_{j^{'}} a_{j^{'}}^i   \\
			0 & & \text{otherwise}  
		\end{aligned}
		\right.
\end{equation}

 Following Garg et al.~\cite{garg2019jointly}, we utilize the attentional probability of the penultimate layer, i.e., $l = L-1$, as the alignment result, which is shown to provide the best alignment result in previous studies~\cite{garg2019jointly}. Afterward, we group together the vectors $a^{i}$ to obtain the attention matrix $A_{I\times J}$. The importance of tokens in the Fact section is calculated as follows:
\begin{equation}\label{eqn-14} 
X_i = \sum_{j=1}^{J} \Lambda _{i,j}A_{i,j}
\end{equation}

According to $X_i$, we choose the top $p\%$ importance token in the Fact section as the key facts $F_{key}$, and the others are considered as non-key facts $F_{non-key}$. Then, we employ mean pooling on the embedding of these token to get their representations. In an ideal vector space for legal cases, $[CLS]$ embeddings are expected to be consistent with representations of key facts while far from non-key facts. Thus, contrastive learning is employed to achieve this objective, where representations of key facts are treated as positive examples $h_p$ and those of non-key facts are negative examples $h_n$. The loss function $L_{CON}$ is formulated as follows:
\begin{equation}\label{eqn-15} 
  L_{CON} =
-\log_{}{    \frac{exp(sim(h_f,h_p))}{exp(sim(h_f,h_p))+exp(sim(h_f,h_n))}}
\end{equation}
where $sim(\cdot )$ is the dot-product function. In practice, the in-batch negative technique~\cite{karpukhin2020dense} is employed to better utilize positive and negative samples from the same batch.
To be specific, given a case $q$, both positive and negative samples of the other cases in the same batch are considered as negative samples for case $q$. Notably, all positive and negative examples are obtained unsupervised without any manual annotation. This unsupervised text-level contrastive learning loss can effectively warm up the embedding space and improve the discriminative ability of the DELTA.

\subsection{Training Objective}

Finally, the optimization objective of the model is the combination of the above losses, which is formulated as: 
\begin{equation}\label{eqn-16} 
  L_{Total} = L_{MLM} + L_{REA} +  L_{DEC} + \lambda_1 L_{TLM} + \lambda_2 L_{CON} 
\end{equation}
where $\lambda_1$ and $\lambda_2$ is the hyperparameters.

\section{Experiment Setting}
In this section, we provide a comprehensive description of our experimental setup, including research questions, datasets, baseline methods, metrics, and implementation details.

\subsection{Research Questions}
We conducted experiments to investigate the following research questions:

\begin{itemize}[leftmargin=*]
    \item \textbf{RQ1: } How does DELTA perform compared to other baselines under the zero-shot setting?
    \item \textbf{RQ2: } With the same fine-tuning strategy, how does DELTA perform compared to other baselines?
    \item \textbf{RQ3: } What is the impact of different components on the performance of DELTA?
    \item \textbf{RQ4:} How about the discriminative ability of DELTA compared to the previous state-of-the-art models?
\end{itemize}

\subsection{Datasets and Metrics}
In this paper, we conducted experiments on both Chinese and English legal case retrieval benchmarks. Below, we provide detailed descriptions of these datasets:

\begin{itemize}[leftmargin=*]
    \item[-] \textbf{LeCaRD}~\cite{ma2021lecard} is a widely used legal case retrieval dataset under the Chinese legal system. It contains 107 queries and 43,000 candidate case documents.

    \item[-] \textbf{CAIL2022-LCR} is another Chinese legal case retrieval dataset, which has been provided as the test set for the CAIL2022 legal case retrieval competition~\footnote{\url{http://cail.cipsc.org.cn/task3.html?raceID=3&cail_tag=2022}}. This dataset comprises 130 queries and 13,000 candidate cases.
    
    \item[-] \textbf{COLIEE2022}~\cite{kim2022coliee} is an English dataset that serves as the official dataset for COLIEE2022 Task 1.  This dataset consists of two parts: the training set and the test set. The training set comprises 898 queries and 3,531 candidate cases, while the test set includes 300 queries and 1,263 candidate cases.
    
    \item[-] \textbf{COLIEE2023}~\cite{goebel2023summary} serves as the benchmark for evaluating legal case retrieval techniques in the COLIEE2023 competition. This dataset contains 959 query cases against 4,400 candidate cases for training and 319 query cases against 1,335 candidate cases for testing.

\end{itemize}

We follow the official metrics for these datasets to evaluate performance.
Specifically, for LeCaRD and CAIL2022-LCR, we employ Precision@5, Recall@5, F1 score, NDCG@10 and NDCG@30. Furthermore, for the COLIEE datasets, we provide results for Precision@5, Recall@5, F1 score, R@100, and R@500. 
The reported results represent the averages across all test cases.

\begin{table*}[t]
\tabcolsep=3.5pt
\caption{Comparisons between DELTA and various baselines on Chinese benchmark on zero-shot setting. */** denotes that DELTA performs significantly better than baselines at $p < 0.05/0.01$ level using the Fisher randomization test~\cite{rubin1980randomization}. Best method in each column is marked bold. }
\label{zero-shot-chinese}
\begin{tabular}{lllllllllll}
\hline
\multicolumn{1}{l|}{\multirow{2}{*}{Model}} & \multicolumn{5}{c|}{LeCaRD}                                                                                                                                     & \multicolumn{5}{c}{CAIL2022-LCR}                                                                                                                       \\
\multicolumn{1}{l|}{}                       & \multicolumn{1}{c}{Precision} & \multicolumn{1}{c}{Recall} & \multicolumn{1}{c}{F1\_score} & \multicolumn{1}{c}{NDCG@10} & \multicolumn{1}{l|}{NDCG@30}         & \multicolumn{1}{c}{Precision} & \multicolumn{1}{c}{Recall} & \multicolumn{1}{c}{F1\_score} & \multicolumn{1}{c}{NDCG@10} & \multicolumn{1}{c}{NDCG@30} \\ \hline \hline 
\multicolumn{11}{c}{\textbf{Sparse Retrieval Models}}                                                                                                                                                                                                                                                                                                             \\ \hline
\multicolumn{1}{l|}{BM25}                   & 0.8916                        & 0.1748                     & 0.2922                        & 0.7115**                    & \multicolumn{1}{l|}{0.8172*}         & 0.8477**                        & 0.2018**                     & 0.3259**                        & 0.7303**                    & 0.8304*                      \\
\multicolumn{1}{l|}{QLD}                   & 0.8897                        & 0.1737                     & 0.2906                        & 0.7157**                     & \multicolumn{1}{l|}{0.8373}          & 0.8538**             & 0.2004**                     & 0.3246**                        & 0.7535**                      & 0.8545             \\ \hline \hline 
\multicolumn{11}{c}{\textbf{General Pre-trained Models}}                                                                                                                                                                                                                                                                                                               \\ \hline
\multicolumn{1}{l|}{Chinese BERT}           & 0.6654**                      & 0.1263**                   & 0.2123**                      & 0.5252**                    & \multicolumn{1}{l|}{0.5374**}        & 0.6600**                      & 0.1487**                   & 0.2427**                      & 0.5604**                    & 0.5618**                    \\
\multicolumn{1}{l|}{Chinese RoBERTa}        & 0.8841                        & 0.1778                     & 0.2960                        & 0.7438**                      & \multicolumn{1}{l|}{0.7897**}        & 0.9046                       & 0.2180                   & 0.3513                       & 0.8043**                    & 0.8518                    \\
\hline \hline
\multicolumn{11}{c}{\textbf{Dense Retrieval Models}}                                                                                                                                                                                                                                                                                                    \\ \hline
\multicolumn{1}{l|}{Condenser}              & 0.8280**                      & 0.1632**                    & 0.2727**                       & 0.6469**                    & \multicolumn{1}{l|}{0.7125**}        & 0.8092**                      & 0.1892**                   & 0.3067**                      & 0.6925**                    & 0.7402**                    \\
\multicolumn{1}{l|}{coCondenser}            & 0.8411**                      & 0.1648**                    & 0.2756**                       & 0.6719**                    & \multicolumn{1}{l|}{0.7404**}        & 0.8138**                      & 0.1931**                   & 0.3121**                       & 0.7063**                    & 0.7607**                    \\
\multicolumn{1}{l|}{SEED}                   & 0.8411**                      & 0.1575**                   & 0.2653**                      & 0.6721**                    & \multicolumn{1}{l|}{0.7330**}        & 0.8369**                      & 0.1906**                   & 0.3105**                      & 0.7365**                    & 0.7815**                    \\
\multicolumn{1}{l|}{COT-MAE}                & 0.8467**                      & 0.1567**                   & 0.2644**                      & 0.6815**                    & \multicolumn{1}{l|}{0.7089**}        & 0.8446**                      & 0.1996**                   & 0.3229**                      & 0.7274**                    & 0.7311**                    \\
\multicolumn{1}{l|}{RetroMAE}               & 0.8505**                      & 0.1675**                    & 0.2799**                       & 0.6876**                    & \multicolumn{1}{l|}{0.7326**}        & 0.8531**                      & 0.1974**                   & 0.3206**                      & 0.7378**                    & 0.7770**                    \\ \hline \hline
\multicolumn{11}{c}{\textbf{Legal-oriented Pre-trained Models}}   \\ \hline

\multicolumn{1}{l|}{BERT\_xs}               & 0.7159**                      & 0.1377**                   & 0.2309**                      & 0.5695**                    & \multicolumn{1}{l|}{0.5751**}        & 0.6462**                      & 0.1369**                   & 0.2259**                      & 0.5236**                    & 0.5206**                    \\
\multicolumn{1}{l|}{SAILER}                 & 0.9028               & 0.1902            & 0.3142               & 0.7979             & \multicolumn{1}{l|}{0.8485} & 0.8938               & 0.2310            & 0.3671               & 0.8319             & 0.8660             \\

\multicolumn{1}{l|}{Lawformer}              & 0.8056**                      & 0.1552**                   & 0.2603**                      & 0.6216**                    & \multicolumn{1}{l|}{0.6362**}        & 0.7908**                      & 0.1820**                   & 0.2935**                      & 0.6908**                    & 0.6988**                    \\

 \hline
\multicolumn{1}{l|}{DELTA}                 & \textbf{0.9308}               & \textbf{0.1959}            & \textbf{0.3236}               & \textbf{0.8117}             & \multicolumn{1}{l|}{\textbf{0.8579}} & \textbf{0.9077}               & \textbf{0.2377}            & \textbf{0.3767}               & \textbf{0.8379}             & \textbf{0.8709}     \\     
\hline

\end{tabular}
\end{table*}

\begin{table*}[t]
\caption{Comparisons between DELTA and various baselines on English benchmark on zero-shot setting. */** denotes that DELTA performs significantly better than baselines at $p < 0.05/0.01$ level using the Fisher randomization test~\cite{rubin1980randomization}. Best method in each column is marked bold. }
\label{zero-shot-english}
\begin{tabular}{lllllllllll}
\hline
\multicolumn{1}{l|}{\multirow{2}{*}{Model}} & \multicolumn{5}{c|}{COLIEE2022}                                                                              & \multicolumn{5}{c}{COLIEE2023}                                                          \\
\multicolumn{1}{l|}{}                       & Precison        & Recall          & F1\_score       & R@100           & \multicolumn{1}{l|}{R@500}           & Precison        & Recall          & F1\_score       & R@100           & R@500           \\ \hline \hline

\multicolumn{11}{c}{\textbf{General Pre-trained Models}}                                                                                                                                                                                             \\ \hline
\multicolumn{1}{l|}{BERT}                   & 0.0340**        & 0.0403**        & 0.0369**        & 0.2811**        & \multicolumn{1}{l|}{0.6390**}        & 0.0338**        & 0.0628**        & 0.0440**        & 0.3138**        & 0.6909**        \\
\multicolumn{1}{l|}{RoBERTa}                & 0.0506**        & 0.0602**        & 0.0550**        & 0.3223**        & \multicolumn{1}{l|}{0.6596**}        & 0.0357**        & 0.0663**        & 0.0464**        & 0.3214**        & 0.6926**        \\ \hline \hline
\multicolumn{11}{c}{\textbf{Dense Retrieval Models}}                                                                                                                                                                                                 \\ \hline
\multicolumn{1}{l|}{Condenser}              & 0.0367**        & 0.0435**        & 0.0398**        & 0.2813**        & \multicolumn{1}{l|}{0.6344**}        & 0.0338**        & 0.0628**        & 0.0440**        & 0.3066**        & 0.6888**        \\
\multicolumn{1}{l|}{coCondenser}            & 0.0853          & 0.1013          & 0.0926          & 0.4778*         & \multicolumn{1}{l|}{0.7906*}         & 0.0744          & 0.1385          & 0.0969          & 0.5377**        & 0.8751          \\
\multicolumn{1}{l|}{SEED}                   & 0.0526**        & 0.0625**        & 0.0571**        & 0.3066**        & \multicolumn{1}{l|}{0.6448**}        & 0.0369**        & 0.0686**        & 0.0480**        & 0.3538**        & 0.7327**        \\
\multicolumn{1}{l|}{COT-MAE}                & 0.0493**        & 0.0584**        & 0.0535**        & 0.3011**        & \multicolumn{1}{l|}{0.6249**}        & 0.0407**        & 0.0756**        & 0.0529**        & 0.3663**        & 0.7205**        \\
\multicolumn{1}{l|}{RetroMAE}               & 0.0945          & 0.1016          & 0.0957          & 0.4908          & \multicolumn{1}{l|}{0.8061}          & 0.0736**        & 0.1378          & 0.0959          & 0.5363**        & 0.8670*         \\ \hline \hline
\multicolumn{11}{c}{\textbf{Legal-oriented Pre-trained Models}}                                                                                                                                                                                      \\ \hline
\multicolumn{1}{l|}{LEGALBERT}              & 0.0193**        & 0.0229**        & 0.0209**        & 0.1666**        & \multicolumn{1}{l|}{0.4902**}        & 0.0163**        & 0.0302**        & 0.0211**        & 0.1715**        & 0.5412**        \\
\multicolumn{1}{l|}{SAILER}                 & 0.0733*         & 0.0870*         & 0.0796*         & 0.4606**        & \multicolumn{1}{l|}{0.7765**}        & 0.0639**        & 0.1187*         & 0.0831*         & 0.4957**        & 0.8391**        \\ \hline
\multicolumn{1}{l|}{DELTA}                  & \textbf{0.0980}          & \textbf{0.1163}          & \textbf{0.1064}          & \textbf{0.5053}          & \multicolumn{1}{l|}{\textbf{0.8137}}          & \textbf{0.0828}          & \textbf{0.1536}          & \textbf{0.1075}          & \textbf{0.5714}          & \textbf{0.8956}          \\ \hline
\end{tabular}
\end{table*}


\begin{table*}[t]
\caption{Comparisons between DELTA and various baselines on English benchmark with fine-tuning setting. */** denotes that DELTA performs significantly better than baselines at $p < 0.05/0.01$ level using the Fisher randomization test~\cite{rubin1980randomization}. Best method in each column is marked bold. }
\vspace{-4mm}
\begin{tabular}{lllllllllll}
\hline
\multicolumn{1}{l|}{\multirow{2}{*}{Model}} & \multicolumn{5}{c|}{COLIEE2022}                                                                              & \multicolumn{5}{c}{COLIEE2023}                                                          \\
\multicolumn{1}{l|}{}                       & Precison        & Recall          & F1\_score       & R@100           & \multicolumn{1}{l|}{R@200}           & Precison        & Recall          & F1\_score       & R@100           & R@500           \\ \hline \hline
\multicolumn{11}{c}{\textbf{Sparse Retrieval Models}}                                                                                                                                                                                                \\ \hline
\multicolumn{1}{l|}{BM25}                   & 0.1307**        & 0.1552**        & 0.1418**        & 0.5866**        & \multicolumn{1}{l|}{0.8416**}        & 0.1222          & 0.2270          & 0.1589          & 0.6612**        & 0.8835**        \\
\multicolumn{1}{l|}{QLD}                    & 0.1313**        & 0.1559**        & 0.1426**        & 0.6326**        & \multicolumn{1}{l|}{0.8804**}        & 0.1191*          & 0.2212*          & 0.1548*          & 0.7113*         & 0.9255*         \\ \hline \hline
\multicolumn{11}{c}{\textbf{General Pre-trained Models}}                                                                                                                                                                                             \\ \hline
\multicolumn{1}{l|}{BERT}                   & 0.1146**        & 0.1362**        & 0.1245**        & 0.5752**        & \multicolumn{1}{l|}{0.8699**}        & 0.0959**        & 0.1781**        & 0.1247**        & 0.6419**        & 0.9264*         \\
\multicolumn{1}{l|}{RoBERTa}                & 0.1524**        & 0.1805**        & 0.1650**        & 0.7517          & \multicolumn{1}{l|}{0.9414}          & 0.1003**        & 0.1862**        & 0.1303**        & 0.6967**        & 0.9070**        \\ \hline \hline
\multicolumn{11}{c}{\textbf{Dense Retrieval Models}}                                                                                                                                                                                                 \\ \hline
\multicolumn{1}{l|}{Condenser}              & 0.2000**        & 0.2375**        & 0.2171**        & 0.7008**        & \multicolumn{1}{l|}{0.9160*}         & 0.1110**        & 0.2061**        & 0.1443**        & 0.7261*         & 0.9386*         \\
\multicolumn{1}{l|}{coCondenser}            & 0.2393**        & 0.2842**        & 0.2598**        & 0.7508          & \multicolumn{1}{l|}{0.9311}          & 0.1197*          & 0.2223*          & 0.1556*          & 0.7355          & 0.9409          \\
\multicolumn{1}{l|}{SEED}                   & 0.2266**        & 0.2692**        & 0.2461**        & 0.7527          & \multicolumn{1}{l|}{0.9391}          & 0.1223          & 0.2270          & 0.1589          & 0.7182*         & 0.9310*         \\
\multicolumn{1}{l|}{COT-MAE}                & 0.2427*         & 0.2882**        & 0.2634**        & 0.7608          & \multicolumn{1}{l|}{0.9412}          & 0.1229          & 0.2282*          & 0.1597          & 0.7347          & 0.9472          \\
\multicolumn{1}{l|}{RetroMAE}               & 0.2206**        & 0.2620**        & 0.2395**        & 0.7396*         & \multicolumn{1}{l|}{0.9264*}         & 0.1072**        & 0.1991**        & 0.1394**        & 0.7431          & 0.9444          \\ \hline \hline
\multicolumn{11}{c}{\textbf{Legal-oriented Pre-trained Models}}                                                                                                                                                                                      \\ \hline
\multicolumn{1}{l|}{LEGALBERT}              & 0.0713**        & 0.0847**        & 0.0774**        & 0.3432**        & \multicolumn{1}{l|}{0.7123**}        & 0.0545**        & 0.1013**        & 0.0709**        & 0.4659**        & 0.8629**        \\
\multicolumn{1}{l|}{SAILER}                 & 0.2540*         & 0.3016*         & 0.2757*         & 0.7364*         & \multicolumn{1}{l|}{0.9325}          & 0.1253          & 0.2328          & 0.1629          & 0.7094**        & 0.9371*         \\ \hline
\multicolumn{1}{l|}{DELTA}                  & \textbf{0.2707} & \textbf{0.3214} & \textbf{0.2938} & \textbf{0.7636} & \multicolumn{1}{l|}{\textbf{0.9493}} & \textbf{0.1316} & \textbf{0.2444} & \textbf{0.1711} & \textbf{0.7493} & \textbf{0.9502} \\ \hline
\end{tabular}
\label{finetune}
\end{table*}

\subsection{Baselines}
We conduct a comparison of DELTA with four distinct categories of baseline models, including Sparse Retrieval Models, General Pre-trained Models, Dense Retrieval Models, and Legal-oriented Pre-trained Models.

Regarding Sparse Retrieval Models, we consider taking BM25~\cite{robertson2009probabilistic} and QLD~\cite{zhai2008statistical} as the baseline models, both of which are classical retrieval models based on lexical matching. Furthermore, General Pre-trained Models include BERT~\cite{devlin2018bert} and RoBERTa~\cite{liu2019roberta}, with adaptations for the Chinese dataset using the corresponding Chinese versions of BERT and RoBERTa. We further extend our comparison to include a range of Dense Retrieval models: Condenser~\cite{gao2021condenser}, coCondneser~\cite{gao2021unsupervised}, SEED~\cite{lu2021less}, COT-MAE~\cite{wu2022contextual}, RetroMAE~\cite{liu2022retromae}. These models have retrieval-oriented optimization objectives and achieve state-of-the-art performance in web search tasks. Finally, we take into account pre-trained language models tailored to legal scenarios. These models comprise:

\begin{itemize}[leftmargin=*]
    \item[-] \textbf{BERT\_xs}~\footnote{\url{http://zoo.thunlp.org/}} is a specialized model designed for criminal law applications. It has been pre-trained on a substantial dataset comprising 663 million Chinese criminal judgments.

    \item[-] \textbf{Lawformer}~\cite{xiao2021lawformer} enhances the attention mechanism to effectively process lengthy legal cases, which is the first Chinese pre-trained language model tailored to legal scenarios.
    
    \item[-] \textbf{LEGAL-BERT}~\cite{chalkidis2020legal} has been developed using a substantial English corpus and shows competitive performance across a diverse range of tasks.

    \item[-] \textbf{SAILER}~\cite{SAILER} is a structure-aware pre-trained model, explicitly tailored for the task of legal case retrieval, and has achieved state-of-the-art results in this domain.
\end{itemize}

The Sparse Retrieval Models are implemented by Pyserini~\footnote{\url{https://github.com/castorini/pyserini}}. As for both General Pre-trained Models and Legal-oriented Pre-trained Models, we utilize the open-source checkpoint directly. For Dense Retrieval Models, we conduct pre-training on the legal corpus with their optimal parameters to ensure equitable comparisons.

\subsection{Implementation Details}

\subsubsection{Pre-training}

We initialize the encoder of the English/Chinese version of DELTA separately from bert-base-uncased/chinese-bert-wwm. The decoders are initialized from scratch. In line with Li et al.~\cite{SAILER}, the Chinese pre-training corpus comprises tens of millions of legal case documents, sourced from China Judgment Online~\footnote{\url{https://wenshu.court.gov.cn/}}. Similarly, the English pre-training corpus comprises an extensive collection of legal case documents from the U.S. federal and state courts~\footnote{\url{https://case.law/}}. For each case, we extract the Fact, Reasoning, and Decision sections using regular matching.

During the pre-training phase, we conduct training for up to 5 epochs using AdamW~\cite{loshchilov2018fixing} optimizer, with a learning rate of 5e-5, and a linear schedule with a warmup ratio of 0.1. The batch size is set to 72. The default mask ratio is set to 0.15 for the encoder and 0.45 for the shallow decoders. The shallow decoder is designed with a single transformer layer, while the deep decoder consists of six transformer layers. Furthermore, the top 60\% of the tokens in the fact section are identified as key facts. The hyperparameters $\lambda_1$ and $\lambda_2$ are both set to 1.

\subsubsection{Fine-tuning}

During the fine-tuning phase, we discard all decoders and solely fine-tune the encoder. The training is performed on the COLIEE training set, employing a contrastive learning loss.
For each given query, we employ BM25 to retrieve the top 100 related documents from the entire corpus, where irrelevant documents are treated as negative examples. The ratio of positive to negative examples is maintained at 1:15. We fine-tune the model up to 20 epochs employing the AdamW~\cite{loshchilov2018fixing} optimizer, with a learning rate set at 5e-6, a batch size of 4, and a linear schedule that includes a warmup ratio of 0.1. All experiments presented in this paper are conducted on 8 NVIDIA Tesla A100 GPUs. To facilitate the reproductivity of our results, we will release the source code for our experiments after the reviewing phase.

\begin{table}[t]
\caption{Ablation study on COLIEE2023 under zero-shot setting. Best results are marked bold.}
\label{ablation}
\begin{tabular}{llll}
\hline
Method         & Precision       & Recall          & F1\_score       \\ \hline
DELTA          & \textbf{0.0828} & \textbf{0.1526} & \textbf{0.1075} \\
w/o $L_{CON}$              & 0.0639                         & 0.1187          & 0.0831          \\
w/o $L_{TLM}$              & 0.0557                         & 0.1036          & 0.0725          \\
w/o $L_{REA}$              & 0.0633                         & 0.1175          & 0.0823          \\
w/o $L_{DEC}$              & 0.0608                         & 0.1129          & 0.0790          \\
w/o $L_{MLM}$              & 0.0796                         & 0.1478          & 0.1035          \\
w/o All                     & 0.0338                         & 0.0628          & 0.0440          \\ \hline
\end{tabular}
\end{table}

\section{Experiment Result}
In this section, we present the experimental results to illustrate the effectiveness of the DELTA.

\subsection{Zero-Shot Evaluation}
To answer \textbf{RQ1}, we conduct zero-shot experiments on four legal benchmarks. The performance comparisons on two Chinese criminal law datasets are shown in Table \ref{zero-shot-chinese}. Based on the results in this table, we can derive the following conclusions:

\begin{itemize}[leftmargin=*]
    \item Without the guidance of training data, BM25 and QLD show competitive performance in legal case retrieval task.
    \item General Pre-trained Models usually show suboptimal performance on legal case retrieval.
    This can be attributed to that their pre-training objectives are not tailored for similarity-based tasks, especially for legal case retrieval.
    \item 
    Dense Retrieval model enhances their ability for similarity modeling, typically resulting in superior performance compared to General Pre-trained Models. However, they still fall short in the legal domain due to a lack of understanding of legal cases.
    \item Legal-oriented Pre-trained Models are generally trained on extensive legal texts, which contributes to a better understanding of legal cases. However, the pre-training objectives of both BERT\_xs and Lawformer do not focus on retrieval tasks, which limits their performance. SAILER, on the other hand, performs better than other pre-trained models, indicating that utilizing structural information in legal cases can lead to more effective case representations.
    \item Finally, We can observe that DELTA achieves the best performance on all metrics. 
    Furthermore, it's worth noting that DELTA doesn't show a significant improvement compared to SAILER. This observation might be attributed to the nature of the Chinese legal system where cases with similar causes are typically considered relevant, rendering the need for discriminative ability less crucial. Nevertheless, by enhancing its discriminative capability, DELTA reaches the state-of-the-art in both Chinese datasets.

\end{itemize}

Table ~\ref{zero-shot-english} presents the performance of DELTA and baselines on the COLIEE2022 and COLIEE2023. From the results, we have the following observations. Among all the pre-trained models, DELTA achieves the best zero-shot results. When compared to SAILER, DELTA achieves a significant improvement, highlighting the critical role of discriminative ability in the determination of case relevance. Without any supervised data, DELTA consistently achieves outstanding retrieval performance across both English and Chinese datasets, which demonstrates the high potential of DELTA in real-world applications.



\subsection{Fine-tuning Evaluation}
To answer \textbf{RQ2}, we further compare DELTA with baselines on English benchmarks. For a fair comparison, we employ the same set of hyperparameters and fine-tuning data across various pre-trained models. As shown in Table ~\ref{finetune}, we have the following findings:
\begin{itemize}[leftmargin=*]
    \item With the guidance of annotated data, the performance of pre-trained models is further improved. However, the majority of these models still face challenges in achieving satisfactory performance.
    \item In comparison to COLIEE2023, pre-trained models a more substantial improvement on the COLIEE2022 dataset. This improvement could be attributed to the greater similarity between the training and testing sets in COLIEE 2022.
    \item Overall, DELTA consistently achieves the best results on both datasets under fine-tuned evaluation. This indicates that the advantages of DELTA are universal, irrespective of the availability of annotated training data.
\end{itemize}

\begin{table}[t]
\caption{The impact of deep decoder layers number on COLIEE2023 under zero-shot setting. Best results are marked bold.}
\small
\begin{tabular}{lllllll}
\hline
Decoder Layer & 2      & 3      & 4      & 5      & 6               & 7      \\ \hline
Precision     & 0.0645 & 0.0727 & 0.0771 & 0.0783 & \textbf{0.0828} & 0.0802 \\
Recall        & 0.1199 & 0.1350 & 0.1431 & 0.1455 & \textbf{0.1526} & 0.1490 \\
F1\_score     & 0.0839 & 0.0945 & 0.1002 & 0.1018 & \textbf{0.1075} & 0.1043 \\ \hline
\end{tabular}
\label{layer}
\end{table}

\begin{table}[t]
\caption{The impact of the ratio of key facts on COLIEE2023 under zero-shot setting. Best results are marked bold.}
\begin{tabular}{llll}
\hline
\textbf{p} & Precision       & Recall          & F1\_score       \\ \hline
10         & 0.0307          & 0.0570          & 0.0399          \\
20         & 0.0501          & 0.0931          & 0.0651          \\
30         & 0.0570          & 0.1059          & 0.0741          \\
40         & 0.0658          & 0.1222          & 0.0855          \\
50         & 0.07398         & 0.1373          & 0.0961          \\
60         & \textbf{0.0828} & \textbf{0.1526} & \textbf{0.1075} \\
70         & 0.0752          & 0.1397          & 0.0977          \\
80         & 0.0708          & 0.1315          & 0.0920          \\ \hline
\end{tabular}
\label{keyfact}
\end{table}

\subsection{Ablation Studies}

To answer \textbf{RQ3}, we conduct an ablation study on the COLIEE 2023 dataset in a zero-shot setting. The experimental results are presented in Table ~\ref{ablation}. From these results, we observe: (1) The removal of the $L_{CON}$ component results in a significant degradation of performance, suggesting that the proposed contrastive learning loss plays a crucial role in learning discriminative representations.
(2) The absence of $L_{TLM}$, which is crucial for training the deep decoder to provide accurate alignment, leads to a significant drop in performance. This highlights the importance of key facts in determining the relevance of legal cases.
(3) Removal of either $L\_{DEC}$ or $L\_{REA}$ also results in a decrease in model performance. This suggests that shallow decoders can create information bottlenecks, enhancing the representational capabilities of text vectors. This finding aligns with observations reported by Li et al.~\cite{SAILER}. (4) Consistent with previous research, the MLM task enhances the model's text comprehension, further boosting performance.  These results demonstrate the effectiveness of our pre-training objectives. Both the representation and discriminative abilities of vectors are crucial for effective legal case retrieval.

\subsection{Hyperparameter Analysis}

In this section, we further investigate the impact of various components within the deep decoder. All reported results here are derived from experiments conducted on the COLIEE 2023 dataset in the zero-shot setting.
 
\subsubsection{Impact of deep decoder layers number}
We initially study the influence of the number of deep decoder layers on performance. As shown in Table ~\ref{layer}, performance consistently improves as the number of deep decoder layers increases, up to a point of 6 layers. Notably, a significant decline in performance is observed when the layer count is reduced to 2. 
We suspect that it is necessary for a certain depth in the decoder to more effectively execute the translation task. Overall, DELTA's performance in relation to its decoder layers exhibits notable robustness.

\subsubsection{Impact of the ratio of key facts}
Subsequently, we explore the effect of the key fact ratio on performance. Specifically, this experiment involves grid searching the parameter $p$ from 10\% to 80\%, in increments of 10\%. 
As shown in Table ~\ref{keyfact}. It is evident that the key fact ratio has a significant impact on model performance.  At lower $p$ values (e.g., 10\%, 20\%), the selected tokens fail to adequately represent the entire case, leading to reduced performance. Conversely, excessively high  $p$ values (e.g., 70\%, 80\%) may include non-key facts, potentially damaging performance. Overall, DELTA maintains commendable performance across a wide range of key fact ratios, with optimal performance achieved at a $p$ value of 60\%.

\begin{figure}[t]
\centering
\includegraphics[width=0.9\columnwidth]{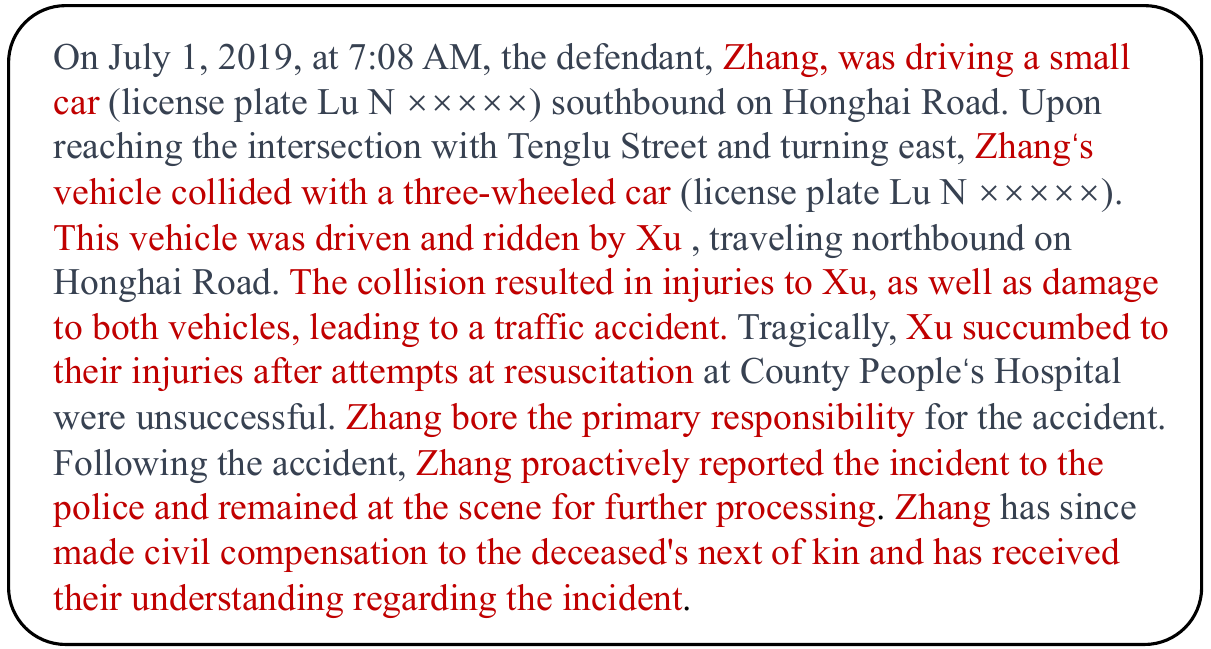}
\caption{An example showing the key facts and non-key facts determined by DELTA. The words in red are key facts, and black are non-key facts.}
\label{example}
\end{figure}

\subsection{Case Study}

In this section, we present a case study to clearly demonstrate the effectiveness of DELTA. As depicted in Figure \ref{example}, the word alignment mechanism within DELTA distinguishes between key and non-key facts. It is evident that crucial elements influencing sentencing, such as surrender and civil compensation, are identified as key facts. These are the aspects that legal professionals focus on and scrutinize in their analysis. Conversely, non-key facts typically encompass basic details like time and location, which are often considered irrelevant noise in the context of case relevance. This case study demonstrates how DELTA efficiently discerns key facts, thereby augmenting the interpretability of legal case searches.

\subsection{Visual Analysis}
\begin{figure}[t]
\vspace{-4mm}
	\centering
	
	\subfigure[SAILER]{
	
	\centering
	\includegraphics[width=0.8\linewidth]{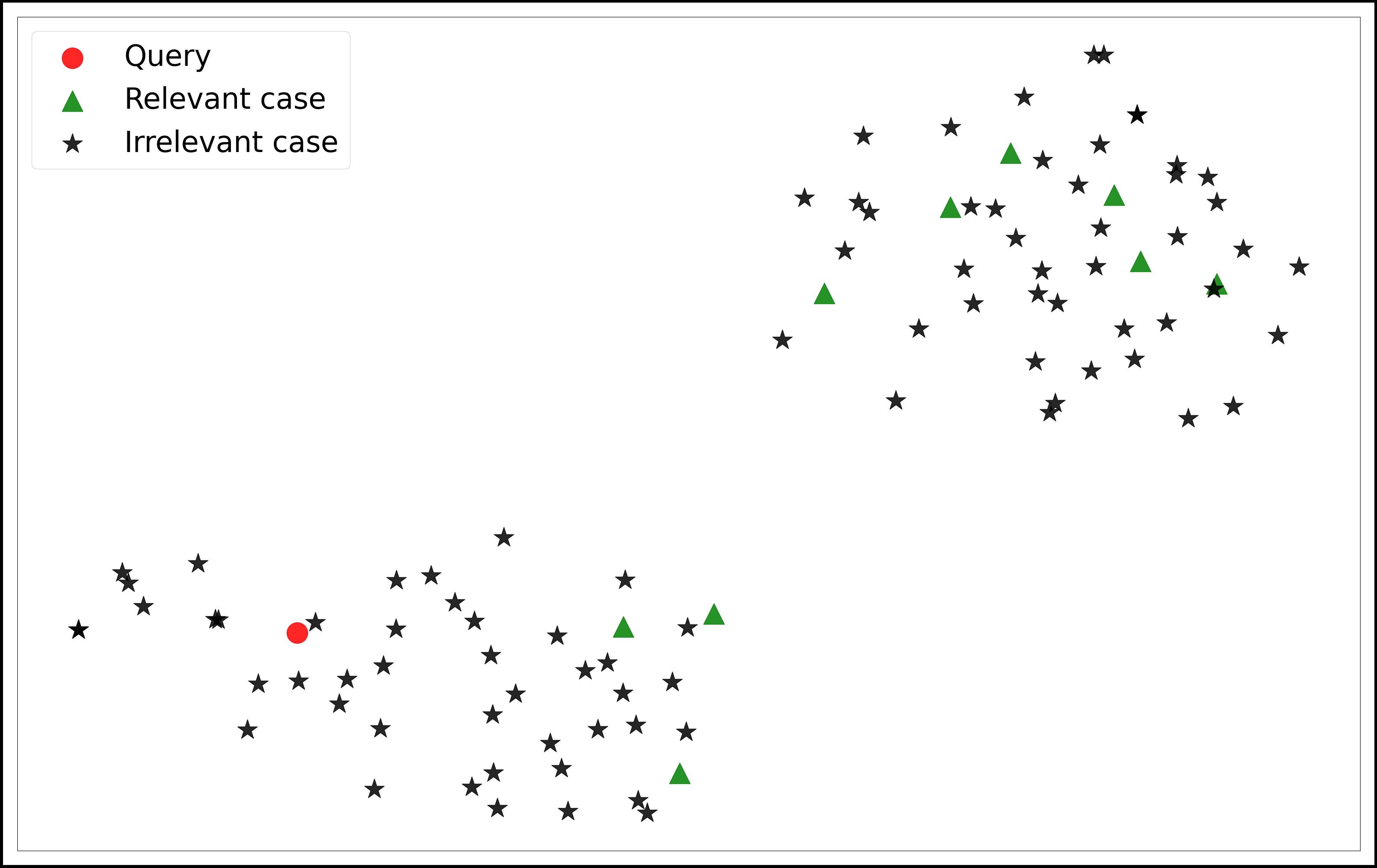}
	
	}
    
	\subfigure[DELTA]{
		
			\centering
			\includegraphics[width=0.8\linewidth]{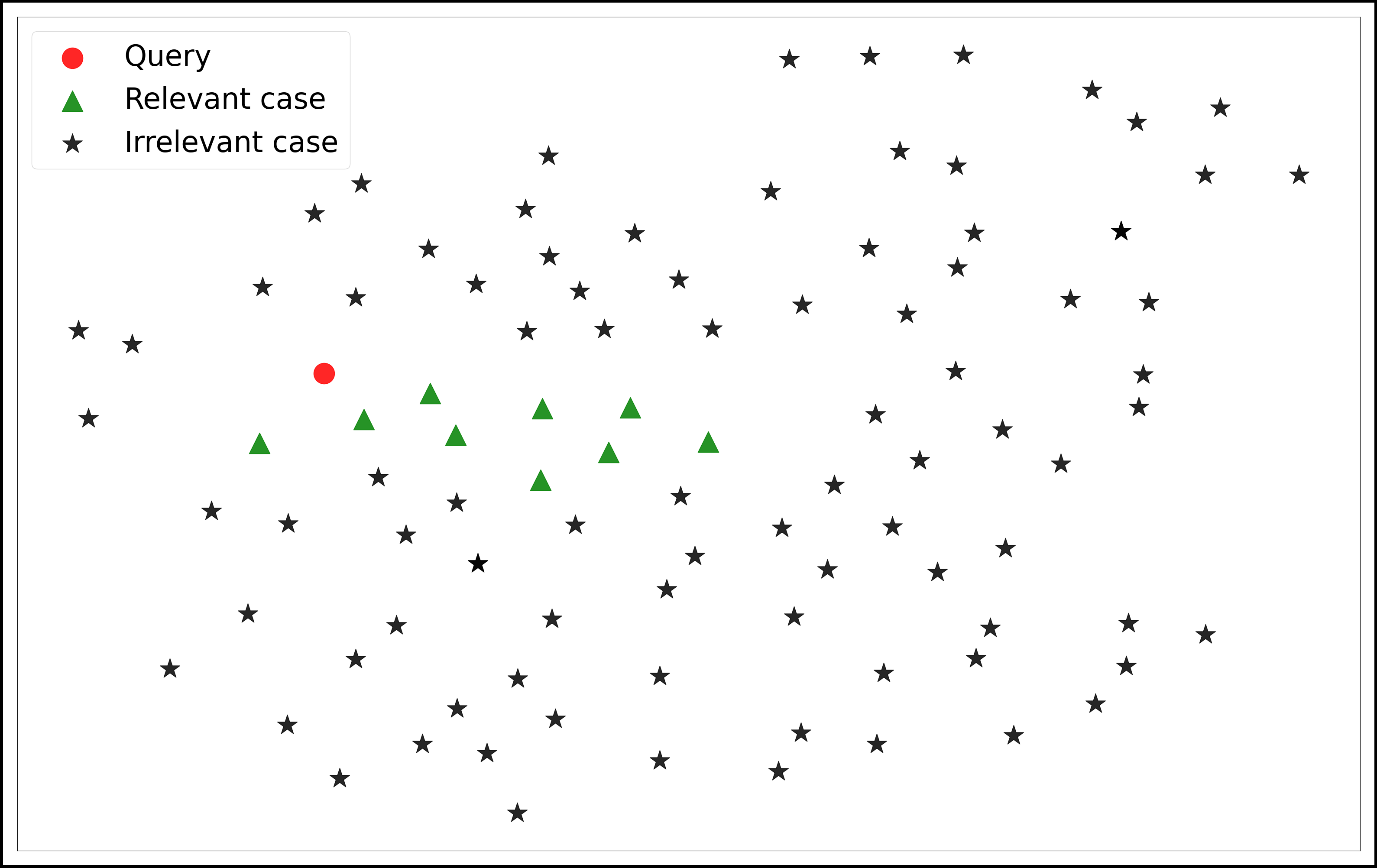}
		
	}

        \vspace{-5mm}
	\caption{Visual analysis of SAILER and DELTA in zero-shot setting.}
	\label{vis}
\vspace{-5mm}
\label{tSNE}
\end{figure}

To answer \textbf{RQ4}, we employed t-SNE to visualize the vector distribution of legal case documents.
This analysis was conducted on the COLIEE 2023 dataset in the zero-shot setting. Specifically, we visualized the sampled query and its top 200 candidate cases. Figure ~\ref{tSNE} displays the results for both SAILER and DELTA.
For SAILER, the distribution of positive samples appears almost random, and the vector distribution is concentrated in a specific region. This is attributed to SAILER focus on modeling the representation of individual cases without considering the relationships between them.
In contrast, DELTA addresses this limitation by utilizing the Structural Word Alignment task to warm up the vector space. As a result, in the vector space of DELTA, positive cases are more closely aligned with the query, and the overall vector distribution is more uniform.
This approach allows DELTA's vector representation to more accurately reflect the relevance between legal cases. Overall, compared to SAILER, DELTA demonstrates a superior ability to generate discriminative case vectors.

\section{Conclusion}
In this paper, we present a novel pre-training framework DELTA for legal case retrieval.
DELTA skillfully utilizes the translation process between different structures of legal case documents to identify key facts. These key facts are further employed to warm up the vector space, aiming to improve the discriminative ability of textual representations.
Experimental results show that our pre-trained objectives contribute significantly to effective retrieval performance. Our method achieves state-of-the-art results on publicly available Chinese and English benchmarks. In the future, we will explore more ways to inject legal knowledge into pre-trained models for deeper understanding and analysis of legal case documents.

\balance
\bibliographystyle{ACM-Reference-Format}
\bibliography{sample-base.bib}
\end{document}